\documentclass[aps,prl,reprint]{revtex4-1}

\pdfoutput=1 

\usepackage{graphicx}							
\usepackage{hyperref}
\usepackage{amssymb}
\usepackage{amsmath}
\usepackage{bm}
\usepackage{bbold}	
\usepackage{braket}
\usepackage{mathtools}

\newcommand{\comment}[1]{}			
\DeclareMathOperator{\tr}{tr}
\DeclareMathOperator{\Tr}{Tr}

\renewcommand{\vec}{\mathbf}

\begin{document}

\title{Cavity superconductor-polaritons}

\author{Andrew A. Allocca} \email{aallocca@umd.edu}
\author{Zachary M. Raines}
\author{Jonathan B. Curtis}
\author{Victor M. Galitski}
\affiliation{Joint Quantum Institute and Condensed Matter Theory Center, University of Maryland, College Park, Maryland, 20742, USA}
\date{\today}

\begin{abstract}
Following the recent success of realizing exciton-polariton condensates in cavities, we examine the hybridization of cavity photons with the closest analog of excitons within a superconductor, states called Bardasis-Schrieffer (BS) modes. 
Though BS modes do not typically couple directly to light, one can engineer a coupling with an externally imposed supercurrent, leading to the formation of hybridized Bardasis-Schrieffer-polariton states, which we obtain both via direct solution and through the derivation of an effective Hamiltonian picture for the model.
These new excitations have nontrivial overlap with both the original photon states and $d$-wave superconducting fluctuations, implying that their condensation could produce a finite $d$-wave component of the superconducting order parameter -- an $s\pm id$ superconducting state.
\end{abstract}
\maketitle

Strong light-matter interaction has been a field of continuing interest for many years~\cite{*[{For a detailed history see e.g. }] [{}] Carusotto2013}.
Formed from the strong coupling of photons in an electromagnetic microcavity and excitons within a semiconductor, exciton-polaritons~\cite{Hopfield1958} and their condensation at high temperatures are by now a well-established experimental milestone~\cite{Weisbuch1992,Wertz2010,Li2013,Sun2017}. 
These objects have generated continued interest in such applications as quantum simulation of solid state physics~\cite{Kavokin2005,Leyder2007,Jacqmin2014,Whittaker2018}, acoustic black hole physics~\cite{Nguyen2015}, and the study of topological properties of quasicrystal states~\cite{Baboux2017}.

Similar cavity hybridization schemes have been proposed within the context of superconducting systems in order to affect the condensation of Cooper-pairs and enhance the strength of superconductivity. 
Specially tailored electromagnetic cavities have been proposed to enhance superconductivity through various mechanisms ~\cite{Laussy2010,Baskaran2012,Cotlet2016,Curtis2018,Sentef2018,Schlawin2018}.
And though there is a rough similarity between the semiconducting and superconducting quasiparticle spectra, both featuring a gap, the superconducting ground state is more complicated than in a semiconductor and therefore the question of how to form polaritons requires more careful consideration. 

The existence of internal exciton-like excitations of the superconducting order parameter was originally proposed by Bardasis and Schrieffer not long after the development of the BCS theory of superconductivity ~\cite{Bardasis1961}. 
These modes, now named Bardasis-Schrieffer (BS) modes, can be thought of as the excitation of Cooper-pairs into states with higher angular momentum than their ground state. 
More precisely, BS modes are gapped, undamped, in-gap fluctuations of the superconducting order parameter in a subdominant pairing channel with a $U(1)$ phase of $\pi/2$ relative to the ground state condensate.
Typically $d$-wave fluctuations are considered about an $s$-wave state, as we will consider here.
These modes have long been sought experimentally but are difficult to detect because they do not directly couple to electromagnetism; their detection has only been recently reported through Raman spectroscopy in iron-based materials~\cite{Kretzschmar2013,Bohm2014,Thorsmolle2016,Jost2018}.

\begin{figure}[t]
    \centering
    \includegraphics[width=\linewidth]{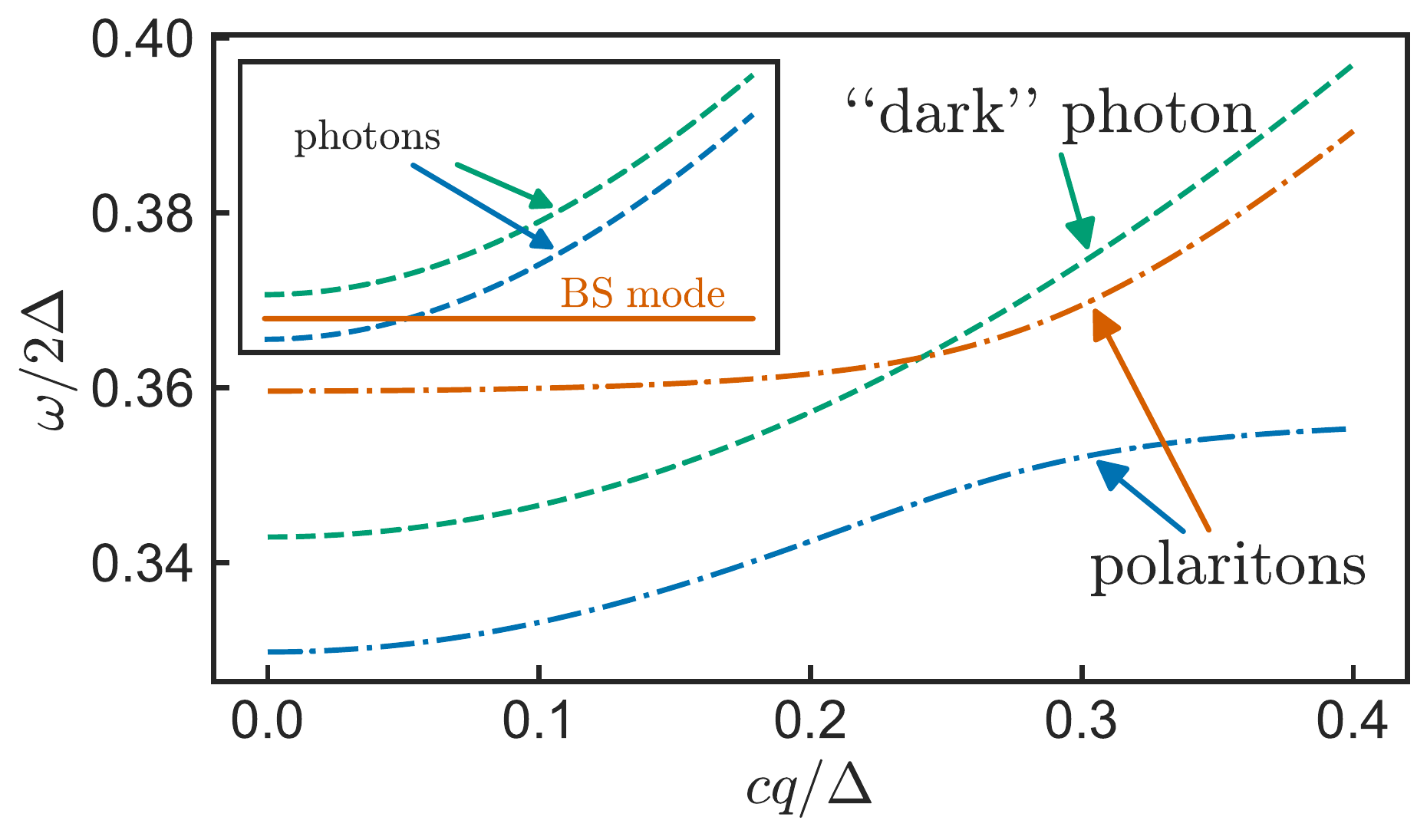}
    \vspace{-7mm}
    \caption{(Color online) The dispersion of the BS-polariton modes (dot-dashed), calculated both numerically and with a simplified analytic method~\cite{Supplement} -- the two give visually identical results.
    Both polariton branches are formed by significant hybridization between photons and the Bardasis-Schrieffer collective mode of the superconductor.
    The dotted line is a ``dark'' photon mode that is always decoupled from the superconductor.
    The strength of the hybridization, generated via a supercurrent, can be tuned by changing the angle $\theta_S$ between the supercurrent and the axis defined by the $d$-wave form factor $f_d$.
    In the main figure the angle is chosen to maximize hybridization, $\theta_S=0$. 
    \textit{Inset} --- the two photon modes (dashed) and the BS mode (solid) when the supercurrent is along a node of $f_d$.
    For such angles there is no hybridization.
    The axes are the same as in the main figure.
    \label{fig:eigenvalues}}
\end{figure}

In our proposed model the BS mode can be hybridized with photons in an appropriately tuned cavity to form polariton states in analogy with the theory of exciton-polariton formation in a semiconductor.
Importantly we show how the lack of a direct coupling to light, which would normally prohibit this hybridization, can be overcome by driving a supercurrent, a method which has similarly been proposed for directly driving the Higgs mode with light~\cite{Moor2017}.
Our main results, presented in Fig.~\ref{fig:eigenvalues}, demonstrate the hybrid Bardasis-Schrieffer polariton dispersions calculated in our model, which can be obtained from an intuitive Hamiltonian model of the coupled modes.
When populated by driving the appropriate photon mode, these polaritons may then condense, producing a finite $d$-wave component of the otherwise purely $s$-wave superconducting order parameter.
Because of the relative phase between BS modes and the $s$-wave condensate, this yields an $s\pm id$ superconducting state.

In what follows we derive the polariton modes from a microscopic model of coupled fermions and cavity photons, and show that to an excellent degree of approximation the results can be described by an effective Hamiltonian for coupled bosonic modes.

We envision a setup consisting of a two-dimensional electron system inside a perfectly reflecting parallel mirror QED cavity.
The 2D electron system is described by a single band fermion action with a BCS interaction decomposed in angular momentum channels. 
With $\hbar=1$ it is
\begin{equation}
    S_\psi = \sum_{k,\sigma}\bar{\psi}_{k,\sigma}\left(-i\epsilon_n + \xi_k\right)\psi_{k,\sigma} - \frac{1}{\beta}\sum_q\sum_{\ell=s,d}g_\ell \bar{\phi}^\ell_q\phi^\ell_q,
\end{equation}
with $\xi_k = k^2/2m-\mu$ the energy with respect to the Fermi surface, $\sigma$ labelling spin, subscript $k$ and $q$ representing both momentum and Matsubara frequency, $g_\ell$ the strength of the interaction in the $\ell$-channel, and the interaction written in terms of the bilinears,
\begin{equation}
    \phi^\ell_q = \sum_k f_\ell(\phi_k) \psi_{-k+\tfrac{q}{2},\downarrow} \psi_{k+\tfrac{q}{2},\uparrow}.
\end{equation}
Importantly, following Bardasis and Schrieffer~\cite{Bardasis1961} we assume the interaction is sizable in both $s$-wave and $d$-wave channels, but a stronger $s$-wave component, $g_s > g_d$, leads to a purely $s$-wave superconducting ground state. 
The form factors are taken to be $f_s(\phi_k) = 1$ and $f_d(\phi_k) = \sqrt{2}\cos(2\phi_k)$.
This choice of $d$-wave form factor breaks the model's full rotational symmetry by choosing an explicit reference axis from which the angle of $\vec{k}$, here called $\phi_k$, is measured, which we expect to be chosen by the underlying crystal structure of the system --- not explicitly present in our continuum model.
The interaction can be straightforwardly decoupled in both angular momentum channels simultaneously with a Hubbard-Stratonovich transformation
\begin{multline}
    S = \sum_k \bar{\Psi}_k \left(-i\epsilon_n \hat{\tau}_0 + \xi_k \hat{\tau}_3\right)\Psi_k + \frac{1}{\beta}\sum_{q,\ell}\frac{1}{g_{\ell}}\vert\Delta^\ell_q\vert^2 \\
    -\frac{1}{\beta}\sum_{k,q}\bar{\Psi}_{k+\tfrac{q}{2}}\sum_\ell f_\ell(\phi_k)\begin{pmatrix} 0 & \Delta^\ell_q \\ \bar{\Delta}^\ell_{-q} & 0 \end{pmatrix}\Psi_{k-\tfrac{q}{2}},
\end{multline}
where we write the result using Nambu spinors $\Psi_k = (\psi_{k,\uparrow},\bar{\psi}_{-k,\downarrow})$, $\hat{\tau}_i$ are the Pauli matrices in Nambu space with $\hat{\tau}_0$ representing the identity, and $\Delta^\ell_q$ are the complex Hubbard-Stratonovich decoupling fields labeled by angular momentum channel.

The cavity is treated as perfectly reflecting boundaries located at $z=0,L$.
The action for photons inside the empty cavity is
\begin{equation}
    S_\text{cav} = -\frac{1}{2\beta}\sum_{q,n,\alpha}A_{\alpha,n,-q}\left[(i\Omega_m)^2 - \omega_{n,q}^2\right]A_{\alpha,n,q}.
\end{equation}
Here $\alpha$ indexes the two cavity polarizations, $n$ labels the quantized modes resulting from the confinement in $z$, and $\omega_{n,q}^2 = \omega_{n,0}^2 + q^2$, with $\omega_{n,0} = n\pi/L$, is the dispersion of photons inside the cavity.
We will consider just the single mode $n=1$, with all others much higher in energy and therefore far from the resonance we will tune to later, so we drop the index from this point onward.
The vector potential is written in terms of polarizations as $\vec{A}_q(z) = \sum_{\alpha} \bm{\epsilon}_{\alpha,\vec{q}}(z)A_{\alpha,q}$, with $\bm{\epsilon}_{\alpha,\vec{q}}(z)$ the polarization vectors inside the cavity~\cite{Supplement}.
We take the electron system to be located in the middle of the cavity, so only the values of these functions at $z=L/2$ must be considered.
Minimal coupling between the cavity photon and the electron system generates the usual terms in the action: a paramagnetic term proportional to $e\vec{v_k}\cdot\vec{A}_q$, with the electron velocity operator $\vec{v_k} = \vec{k}/m$, and a diamagnetic term proportional to $e^2 A_q^2$.
We now drop the diamagnetic term since it is unimportant both in the weak-field regime~\cite{Maissen2014} and for the cavity photon self-energy in the presence of disorder, which is ubiquitous in 2D~\cite{Altland2010,Kamenev2011}.

Our cavity geometry is chosen for calculation simplicity, but in real microwave cavity the transverse nature of the of photon amplitude envelope is more complicated.
The effect of this is to increase the strength of the paramagnetic coupling, an enhancement which we include via a phenomenological enhancement in the light-matter coupling term~\cite{Maissen2014,Bayer2017,Schlawin2018}.

We now consider externally driving a homogeneous supercurrent through the system. 
A supercurrent can be understood as the superconducting condensate moving with respect to the lab frame with constant uniform velocity, with Bogoliubov quasiparticles therefore being defined in the comoving frame, i.e. the supercurrent can be included via a simple Galilean transformation.
Calling the condensate superfluid velocity $\vec{v}_S$, we have $\vec{v_k} \to \vec{v_k} + \vec{v}_S$.
The angle of $\vec{v}_S$ with respect to the axis defined by $f_d(\phi_k)$, as depicted in the inset to Fig.~\ref{fig:theta_S}, is denoted $\theta_S$.
This modifies the quasiparticle dispersion in the lab frame
\begin{equation}
    \xi_k \to \xi_k + \vec{k}\cdot\vec{v}_S + \frac{1}{2}m v_S^2 \equiv \xi_k^S + \vec{k}\cdot\vec{v}_S.
\end{equation}
The term linear in $\vec{k}$ is a Doppler shift in the energy while the one proportional to $v_S^2$ can be absorbed into a (negligible) redefinition of the chemical potential.
The velocity shift also affects the paramagnetic coupling
\begin{equation} \label{eq:paramagnetic}
    S_{\psi-A} \to \frac{X}{\beta}\sum_{k,q}\bar{\Psi}_{k+\tfrac{q}{2}}\underbrace{\left(-e\vec{v_k}\hat{\tau}_0 - e\vec{v}_S\hat{\tau}_3\right) \cdot \vec{A}_q }_{\equiv \hat{\chi}_{k,q}[A]} \Psi_{k-\tfrac{q}{2}}.
\end{equation}
Here $X$ denotes the phenomenological coupling enhancement described above~\cite{Maissen2014,Bayer2017,Schlawin2018}, which we absorb into a redefinition of the charge.
Crucially the Nambu structure for the paramagnetic and supercurrent-induced terms are different, since particle and hole velocities are shifted oppositely,  which is what ultimately allows the coupling of the BS mode to light.
The supercurrent can equivalently be included as a uniform phase winding of $\Delta^s$ which, upon appropriate gauge transformation, reproduces the theory described above while maintaining explicit gauge invariance throughout.

We make the mean-field approximation on the $s$-wave gap function
\begin{multline}
    S = S_{\Delta,s} + S_{\Delta,d} + S_\text{cav} - \sum_k \bar{\Psi}_k \hat{G}^{-1}_k \Psi_k \\
    +\frac{1}{\beta}\sum_{k,q}\bar{\Psi}_{k+\tfrac{q}{2}} \left(\hat{\chi}_{k,q}[A] - \hat{\Delta}^d_{k,q}\right)\Psi_{k-\tfrac{q}{2}},
\end{multline}
with $S_{\Delta,s} = \beta\vert\Delta\vert^2/g_s$ describing the static, homogeneous $s$-wave component $\Delta$, $S_{\Delta,d} = \beta^{-1}\sum_q\vert\Delta^d_q\vert^2/g_d$ describing the $d$-wave fluctuations, $\hat{G}^{-1}_k = (i\epsilon_n - \vec{k}\cdot\vec{v}_S)\hat{\tau}_0 - \xi_k^S\hat{\tau}_3 + \Delta\hat{\tau}_1$ the inverse Nambu Green's function, and 
\begin{equation}
    \hat{\Delta}^d_{k,q} = f_d(\phi_k)\begin{pmatrix} 0 & \Delta^d_q \\ \bar{\Delta}^d_{-q} & 0 \end{pmatrix}.
\end{equation}
We now integrate out the fermions and expand to second order in $\hat{\Delta}^d$ and $\hat{\chi}$.

The mean field value of $\Delta$ is obtained as the saddle point solution in the absence of $\vec{A}$ and $\Delta^d$ but in the presence of the supercurrent,
in keeping with the approximation that $\Delta$ is unaffected by $d$-wave fluctuations and photons.
We are left with
\begin{equation}
    S_\text{eff} = S_d + S_A + S_{d-A},
\end{equation}
with these three terms defined as the parts of the action describing free $d$-wave fluctuations, cavity photons in the presence of the superconducting system, and the supercurrent-generated coupling between them, respectively. 


Since the $d$-wave fluctuations have much greater kinetic mass than photons, we approximate them with a flat dispersion: the value of their energy in the limit $\vec{q}\to 0$.
Additionally, we drop all terms which vanish in the quasiclassical $\xi$-approximation.
Writing $\Delta^d$ in terms of its real and imaginary components as $\Delta^d_q = d^R_q + id^I_q$, with $\bar{d}^{R/I}_q = d^{R/I}_{-q}$, $S_d$ decouples into an action for each component. 
The real mode is within the Bogoliubov quasiparticle continuum, and is therefore overdamped ~\cite{Bardasis1961,Maiti2015}.
It also remains decoupled from photons despite the supercurrent so we do not consider it further.
The imaginary mode, on the other hand, is the in-gap Bardasis-Schrieffer collective mode.
Renaming $d^I_q$ to $d_q$, the BS mode action is
\begin{equation} \label{eq:BSaction}
    S_d = \frac{1}{\beta}\sum_q d_{-q}\left[\frac{1}{g_d} + \sum_\vec{k} f_d(\phi_k)^2\frac{2\lambda_k\,\delta n_\vec{k}}{(i\Omega_m)^2-(2\lambda_k)^2}\right] d_{q},
\end{equation}
where $\lambda_k = \sqrt{\left(\xi^{S}_k\right)^2 + \Delta^2}$ is the quasiparticle energy in the comoving frame, $E^\pm_\vec{k} = \pm\lambda_k + \vec{k}\cdot\vec{v}_S$ is the Doppler-shifted energy, and $\delta n_\vec{k} = n_F(E^-_\vec{k})-n_F(E^+_\vec{k})$, where $n_F$ is the Fermi function.


The photon sector of the action consists of the empty cavity action $S_\text{cav}$ plus a self-energy term due to the superconductor,
\begin{equation} \label{eq:photonaction}
    S_A = -\frac{1}{2\beta}\sum_{q,\alpha,\beta} A_{\alpha,-q}\left[\left((i\Omega_m)^2-\omega_q^2\right)\delta_{\alpha\beta} - \Pi_{\alpha\beta,q}\right]A_{\beta,q}. 
\end{equation}
The matrix $\Pi_{\alpha\beta,q}$ is the electromagnetic linear response function of the superconducting system written in the cavity polarization basis~\cite{Supplement}.


Within the approximations discussed above, in particular the $\vec{q}\to 0$ limit imposed on superconducting fluctuations, the coupling between photons and the BS mode arises \emph{entirely} through the supercurrent-induced term,
\begin{multline} \label{eq:couplingaction}
    S_{d-A} = -\frac{ie\Delta}{\beta} \sum_{\vec{k},q,\alpha} f_d(\phi_k)\frac{i\Omega_m\,\delta n_\vec{k}}{(i\Omega_m)^2 - (2\lambda_k)^2} \frac{\vec{v}_S\cdot\bm{\epsilon}_{\alpha,q}}{\lambda_k} \\
    \times \left(A_{\alpha,q}\, d_{-q} - A_{\alpha,-q}\,d_q\right),
\end{multline}
consistent with the known result that the BS mode does not normally couple linearly to light. 
As a consequence, the BS mode only couples to the component of the vector potential parallel to the supercurrent.

The action is straightforwardly written in terms of a hybrid inverse Green's function
\begin{equation}
    \label{eq:hybridgf}
    S_\text{eff} = \frac{1}{2\beta}\sum_q (d_{-q},A_{\alpha,-q}) \begin{pmatrix} D^{-1}_{\text{BS},q} & g_{\alpha,q}\delta_{\alpha\beta} \\ g_{\alpha,q}^\ast\delta_{\alpha\beta} & D^{-1}_{\alpha\beta,q} \end{pmatrix} \begin{pmatrix} d_q \\ A_{\beta,q} \end{pmatrix},
\end{equation}
with sums over repeated indices and with $D^{-1}_{\text{BS},q}$, $D^{-1}_{\alpha\beta,q}$, and $g_{\alpha,q}$ defined implicitly through Eqs.~\eqref{eq:BSaction}--
\eqref{eq:couplingaction}.
However, a more intuitive description can be obtained by making a harmonic approximation to the BS action and expanding in terms of BS and photon mode operators
\begin{equation}
    d_q = \frac{b_q + \bar{b}_{-q}}{\sqrt{2K\Omega_\text{BS}}} \qquad A_{\alpha,q} = \frac{a_{\alpha,q}+\bar{a}_{\alpha,-q}}{\sqrt{2\omega_q}},
\end{equation}
where the BS frequency $\Omega_\text{BS}$ is defined through $D^{-1}_\text{BS}(\Omega_\text{BS},\vec{q}) = 0$ and $K \equiv \partial^2 D^{-1}_\text{BS}(z,\vec{q})/\partial z^2\vert_{z=\Omega_\text{BS}}$.
In rewriting the action in terms of the mode operators we make the standard approximation of dropping the counter-rotating terms -- an approximation which we verify post-hoc -- and perform a change of basis from photon polarization states to components parallel and perpendicular to the supercurrent.
This allows the action to be written in terms of an effective bosonic Hamiltonian
\begin{equation} 
    S_\text{eff} \approx \frac{1}{\beta}\sum_q \left(\bar{b}_q,\bar{a}^\parallel_q,\bar{a}^\perp_q\right) \left(-i\Omega_m \check{\mathbb{1}} + \check{H}^\text{eff}_\vec{q}\right)
\begin{pmatrix}
b_q \\ a^\parallel_q \\ a^\perp_q
\end{pmatrix}.
\end{equation}
Setting the frequency inside $g$ to the BS frequency and keeping only to lowest order in $\vec{q}$, the effective Hamiltonian~\cite{Supplement}
is
\begin{equation} \label{eq:hamiltonian}
    \check{H}^\text{eff}_q = \begin{pmatrix}
    \Omega_\text{BS} & g_q & 0 \\
    g_q & \omega_q + \Pi^S_q & 0 \\
    0 & 0 & \omega_q
    \end{pmatrix},
\end{equation}
where $q=|\vec{q}|$, $\Pi^S_q$ is a self-energy shift in the photon mode polarized parallel to the supercurrent, which arises from the supercurrent itself, and
\begin{equation}
    g_q = -iev_S\Delta\sqrt{\frac{2\,\Omega_\text{BS}}{L\, K \omega_q}}\sum_\vec{k} \frac{f_d(\phi_k)}{\lambda_k}\frac{\delta n_\vec{k}}{\Omega_\text{BS}^2 - (2\lambda_k)^2}.
\end{equation}
We see that only one photon mode hybridizes with the BS mode. 
This photon mode and the BS mode can be brought into resonance by tuning the parameters of the system, most straightforwardly the cavity size $L$, which allows them to strongly hybridize. 


For all numerical calculations we use material parameters motivated by the pnictide superconductors~\cite{Cvetkovic2013}, where BS modes have been experimentally detected. 
We set the Fermi energy $\epsilon_F = 100$ meV, the single band effective mass $m = 0.7 m_e$, where $m_e$ is the electron mass, and superconducting critical temperature $T_c = 35$ K.
Furthermore, we tune the size of the cavity $L$ so that $\omega_0 = \pi/L = 0.96\, \Omega_\text{BS}(\theta_S = 0)$, putting cavity photons and the BS mode very near resonance, and we set the phenomenological coupling enhancement to a modest $X=10$, although enhancements of $X=10^2$ or greater have been predicted in similar cavity systems~\cite{Maissen2014,Bayer2017,Schlawin2018}.
Finally we set $T = 0.4 T_c$, the approximate temperature for which we find the maximum $g$, and $v_S = 0.9 \Delta\vert_{v_S=0}/k_F$.
Values of $v_S$ larger than this begin to significantly deplete the $s$-wave condensate, which is undesirable. 

\begin{figure}
    \centering
    \includegraphics[width=\linewidth]{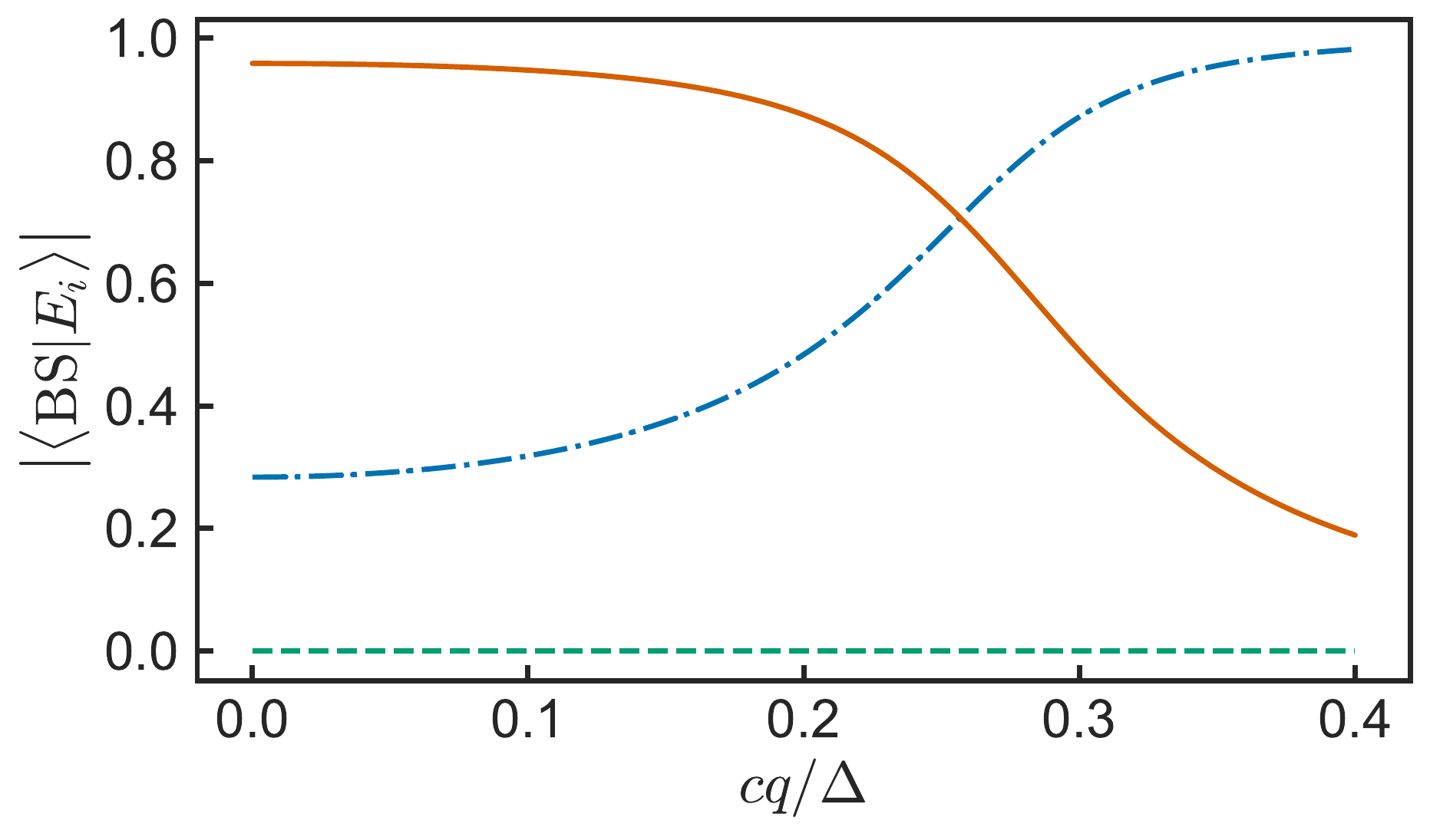}
    \vspace{-7mm}
    \caption{(Color online) The Bardasis-Schrieffer component of the eigenvectors of the effective Hamiltonian Eq.~\eqref{eq:hamiltonian}~\cite{Supplement}.
    The upper (solid) and lower (dot-dashed) polariton modes have significant photon and Bardasis-Schrieffer character showing a strong hybridization between the systems.
    One can also clearly see the ``dark'' photon mode (dashed) which does not hybridize at all with the superconductor's collective mode.
    \label{fig:eigenvectors}}
\end{figure}

\begin{figure}
    \centering
    \includegraphics[width=\linewidth]{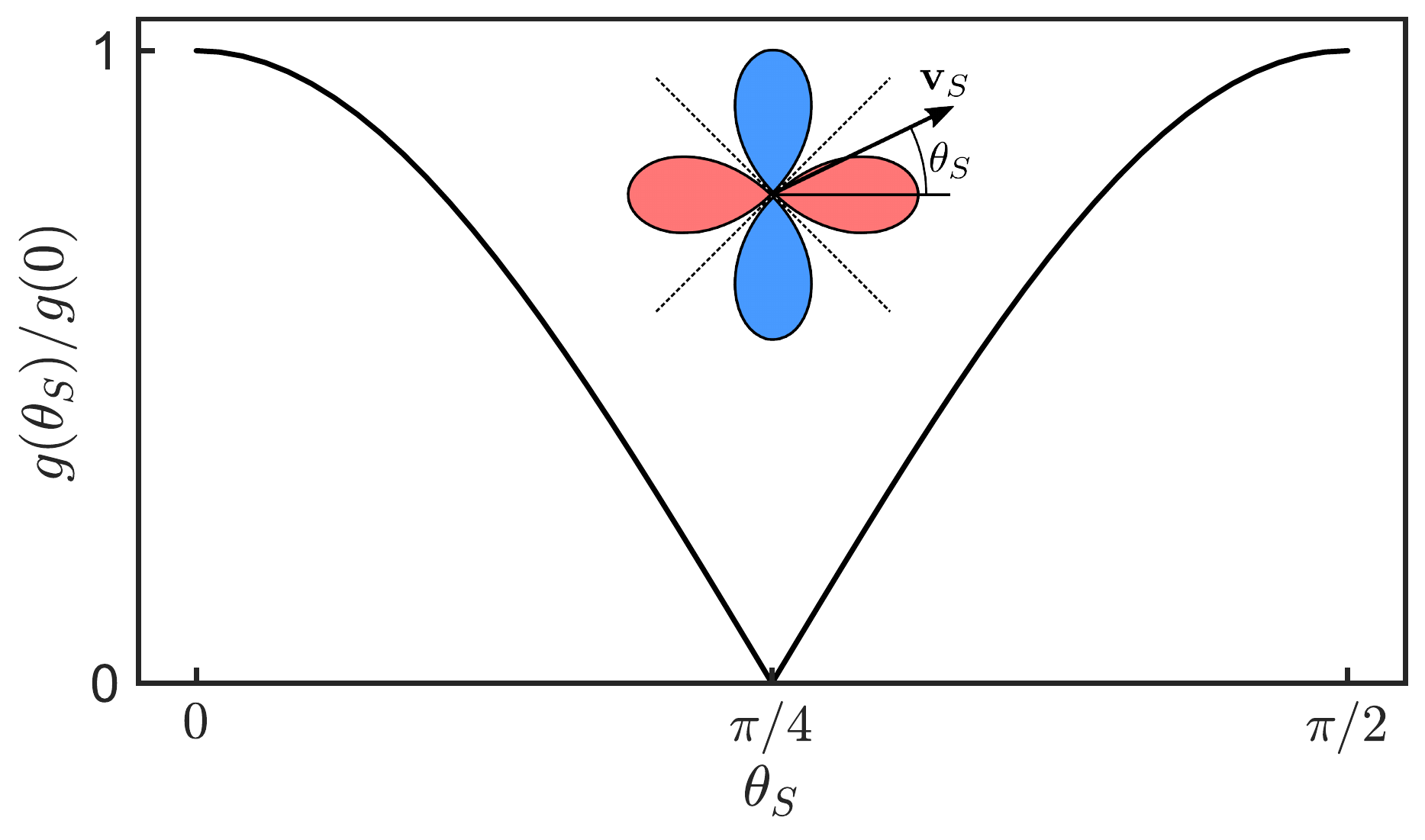}
    \vspace{-7mm}
    \caption{(Color online) The strength of the matrix element $g$, responsible for hybridization, in the effective Hamiltonian \eqref{eq:hamiltonian} as a function of $\theta_S$, the angle between the direction of the supercurrent and the axis defined implicitly by the $d$-wave form factor $f_d(\phi_k) = \sqrt{2}\cos(2\phi_k)$.
    The hybridization is maximal for $\theta_S = m\pi/2$, where $m$ is an integer, while when the supercurrent runs along a node of the $d$-wave form factor, $\theta_S = (2m+1)\pi/4$, the hybridization disappears, as also shown in the inset of Fig.~\ref{fig:eigenvalues}.
    The inset depicts the orientation of the supercurrent with respect to the $d$-wave form factor.
    The color of the lobes represent the relative sign of $f_d$ for different angles, and the dashed lines are the nodes, where the function vanishes.
    \label{fig:theta_S}}
\end{figure}

In order to obtain the polariton modes we both directly solve for the poles of the hybridized Green's function \eqref{eq:hybridgf} and calculate the eigenvalues of the effective Hamiltonian \eqref{eq:hamiltonian}, which can be diagonalized analytically~\cite{Supplement}.
The results of both approaches are in excellent agreement, and the dispersions of the polariton modes are plotted for both methods in Fig.~\ref{fig:eigenvalues}. 
We find that one of the photon modes can be made to strongly hybridize with the BS mode, while the other, the ``dark'' photon, always remains distinct. 
This is made especially clear by examining the BS component of the eigenvectors of the effective Hamiltonian, as shown in Fig.~\ref{fig:eigenvectors}.
The effect is also sensitive to the direction of the supercurrent with respect to the axis defined by the $d$-wave form factor.
Changing this angle $\theta_S$ shifts the frequency of the BS mode relative to the photon modes, as well as dramatically changing the strength of the hybridization itself, as shown in Fig.~\ref{fig:theta_S}. 
The hybridization is strongest when the supercurrent is along one of the maxima of the $d$-wave form factor -- $\theta_S = m\pi/2$, $m\in\mathbb{Z}$ -- and vanishes when the supercurrent is along one of the nodes -- $\theta_S=(2m+1)\pi/4$.

In this work we have shown that driving a supercurrent through a superconductor in a planar microcavity leads to hybridization of cavity photons with the collective modes of the superconductor.
In particular two polaritonic bands form which have significantly mixed character.
This overlap provides a means for observation and control of the normally hard to access Bardasis-Schrieffer mode, and as in the case of exciton-polaritons these dispersions could in principle be measured with $k$-space imaging of the photonic component of the polariton state~\cite{Kasprzak2006}.
Additionally, the condensation observed in exciton-polariton systems~\cite{Wertz2010,Li2013,Sun2017} suggests that proper driving of these superconductor polariton modes could lead to condensation in a subdominant $d$-wave channel and the formation of a non-equilibrium $s\pm id$ superconducting state.
Should such a state be realized experimentally, the nature of the construction allows for tuning of the state \emph{in situ} through the externally applied supercurrent.

\begin{acknowledgments}
We would like to thank Mohammad Hafezi for valuable discussions.
This work was supported by NSF DMR-1613029 (Z.R.), DOE-BES (DESC0001911) and the Simons Foundation (V.G.), US-ARO contract No. W911NF1310172 (A.A.), the National Science Foundation Graduate Research Fellowship Program under Grant No. DGE1322106 (J.C.)
\end{acknowledgments}

\bibliography{references}






\clearpage
\onecolumngrid
\appendix
\begin{center}
{\bf\large Supplement to Cavity Superconductor-Polaritons}
\end{center}
\setcounter{secnumdepth}{2}

\section{Cavity Model}

The model of the photonic sector used in this work is that of a parallel mirror cavity consisting of two conducting plates of infinite extent in the $x-y$ plane and separated by a distance $L$ along the $z$ axis.
In particular, we have chosen to work with the $n=1$ transverse electric and transverse magnetic solutions, which in the Coulomb gauge have polarization vectors
\begin{equation}
    \label{eq:polarization}
    \begin{gathered}
     \bm{\epsilon}_1(\mathbf{q}, z) = i\sqrt{\frac{2}{L}} \sin\left(\frac{\pi z}{L}\right)\hat{\bm{z}} \times \hat{\bm{q}}\\
    \bm{\epsilon}_2(\mathbf{q}, z) = \sqrt{\frac{2}{L}} \frac{1}{\omega_\mathbf{q}}\left(c q\cos\left(\frac{\pi z}{L}\right)\hat{\bm{z}}
    -i \omega_0 \sin\left(\frac{\pi z}{L}\right) \hat{\bm{q}}\right),
    \end{gathered}
\end{equation}
where the $z$ axis is perpendicular to the plane of the superconductor and the momentum $\mathbf{q}$ is in the plane and with energy 
\begin{equation}
\omega^2_\mathbf{q} = \underbrace{\left(\frac{\pi}{L}\right)^2}_{\omega_0^2} + c^2q^2.
\end{equation}

\section{Effective Hamiltonian}
In the remainder of this supplement we detail the two methods we used to solve the problem which lead to effectively identical results as depicted in Fig. 1 of the main text.
First we derive of an effective Hamiltonian, from which we can analytically obtain the polariton dispersions and eigenvalues.
In the next section we discuss the numerical methods used to solve for the polariton dispersion directly from the hybrid inverse Green's function. 
As explained in the main text, we begin with the fermionic mean field model
\begin{equation}
    S = S_{\Delta,s} + S_{\Delta,d} + S_\text{cav} - \sum_k \bar{\Psi}_k \hat{G}^{-1}_k \Psi_k
    +\frac{1}{\beta}\sum_{k,q}\bar{\Psi}_{k+\tfrac{q}{2}} \left(\hat{\chi}_{k,q}[A] - \hat{\tau}_2 f_d(\phi_k)d_q\right)\Psi_{k-\tfrac{q}{2}},
\end{equation}
which has been obtained via Hubbard-Stratonovich decoupling the interaction terms in the Cooper channel and where $\hat G$ is the Nambu Green's function of the $s$-wave state.
Integrating out fermions and keeping to second order in the photon and Bardasis-Schrieffer fields we obtain a description in terms of only bosonic variables
\begin{equation}
    S_\text{eff} = S_d + S_A + S_{d-A},
\end{equation}
where
\begin{subequations}
\begin{gather}
    S_d = \frac{1}{\beta}\sum_q d_{-q}\left[\frac{1}{g_d} + \sum_\vec{k} f_d(\phi_k)^2\frac{2\lambda_k\,\delta n_\vec{k}}{(i\Omega_m)^2-(2\lambda_k)^2}\right] d_{q} \\
    S_A = -\frac{1}{2\beta}\sum_{q,\alpha,\beta} A_{\alpha,-q}\left[\left((i\Omega_m)^2-\omega_q^2\right)\delta_{\alpha\beta} - \Pi_{\alpha\beta,q}\right]A_{\beta,q} \\
    S_{d-A} = -\frac{ie\Delta}{\beta} \sum_{\vec{k},q,\alpha} f_d(\phi_k)\frac{i\Omega_m\,\delta n_\vec{k}}{(i\Omega_m)^2 - (2\lambda_k)^2} \frac{\vec{v}_S\cdot\bm{\epsilon}_{\alpha,q}}{\lambda_k} \left(A_{\alpha,q}\, d_{-q} - A_{\alpha,-q}\,d_q\right),
\end{gather}
\end{subequations}
and $\epsilon_\alpha$ are the polarizations described in Eq.~\eqref{eq:polarization} evaluated at $z=L/2$.

\subsection{Bardasis-Schrieffer Sector}
We begin by rewriting $S_d$ using the mean field equation for the $s$-wave $\Delta$,
\begin{multline}
    S_d = \frac{1}{\beta}\sum_q d_{-q}\left[\frac{1}{g_d} + \sum_\vec{k} f_d(\phi_k)^2\ \frac{\delta n_\vec{k}}{(i\Omega_m)^2-(2\lambda_k)^2}\ 2\lambda_k \underbrace{- \frac{1}{g_s} + \sum_\vec{k} \frac{\delta n_\vec{k}}{2\lambda_k}}_{= 0} \right] d_q \\
    = \frac{1}{\beta}\sum_q d_{-q}\left[\frac{1}{g_d} - \frac{1}{g_s} + \sum_\vec{k} \frac{\delta n_\vec{k}}{2\lambda_k} \left(\frac{(i\Omega_m)^2 + (2\lambda_k)^2(f_d(\phi_k)^2-1)}{(i\Omega_m)^2-(2\lambda_k)^2}\right) \right] d_q \\
    = \frac{1}{\beta}\sum_q d_{-q}\left[\frac{1}{g_d} - \frac{1}{g_s} + (i\Omega_m)^2\sum_\vec{k} \frac{1}{2\lambda_k} \frac{\delta n_\vec{k}}{(i\Omega_m)^2-(2\lambda_k)^2} + \sum_\vec{k} 2\lambda_k \cos(4\phi_k) \frac{\delta n_\vec{k}}{(i\Omega_m)^2-(2\lambda_k)^2} \right] d_q \\
    \equiv -\frac{1}{2\beta}\sum_q d_{-q} D^{-1}_{\text{BS},q} d_q,
\end{multline}
where in the last line we have defined the BS inverse Green's function.
This rewriting regulates the integration and also allows us to straightforwardly parametrize the Bardasis-Schrieffer frequency in terms of the relative strength of the $s$-wave and $d$-wave interactions. 
In order to change to the mode operator basis the inverse Green's function must be rewritten in a harmonic approximation, i.e. expanding to second order in $i\Omega_m$ around the saddle point solution, which we identify as $\Omega_\text{BS}$.
The result of this expansion is
\begin{equation}
    S_d \approx -\frac{K}{2\beta}\sum_q d_{-q} \left((i\Omega_m)^2 - \Omega_\text{BS}^2\right)d_q,
\end{equation}
with the BS frequency satisfying $D^{-1}_\text{BS}(\Omega_\text{BS},\vec{q}) = 0$ and the constant $K \equiv \partial^2 D^{-1}_\text{BS}(z,\vec{q})/\partial z^2\vert_{z=\Omega_\text{BS}}$.
From this form the transformation to mode operators can be performed without further difficulty:
\begin{equation}
    S_d \to S_b = \frac{1}{\beta}\sum_q \bar{b}_q (-i\Omega_m + \Omega_\text{BS})b_q \quad \text{with }\ d_q = \frac{b_q + \bar{b}_{-q}}{\sqrt{2 K \Omega_\text{BS}}}.
\end{equation}

\subsection{Photon Sector}
The self-energy part of the photon action arises from integrating out the fermions inside the superconductor,
\begin{equation}
    S_\Pi = \frac{1}{2} \Tr\left(\hat{G}\hat{\chi}\hat{G}\hat{\chi}\right) \equiv \frac{1}{2\beta}\sum_q \vec{A}_{-q}\hat{\Pi}_q\vec{A}_q = \frac{1}{2\beta}\sum_{q,\alpha,\beta} A_{\alpha,-q} \Pi_{\alpha\beta,q} A_{\beta,q}.
\end{equation}
In the last equality, reproducing the term in the action above, the response function $\Pi$ has been rotated to the basis of cavity polarizations from the original Cartesian basis of the vector potential,
\begin{equation}
    \Pi_{\alpha\beta,q} = \sum_{i,j} \epsilon^i_{\alpha,-\vec{q}} \Pi^{ij}_q \epsilon^{j}_{\beta,\vec{q}}.
\end{equation}
Though the polarization basis is useful for the change to mode operators, an appropriately chosen Cartesian basis is far better for the evaluation of the $\hat{\Pi}$. 
We choose this basis to be defined as the directions parallel and perpendicular to the axis of the supercurrent because we know that this is the basis most relevant for the hybridization problem; only the component of $\vec{A}_q$ parallel to the supercurrent hybridizes with the BS mode. 

The form of $\hat{\Pi}$ can be extracted from the trace above,
\begin{equation}
    \Pi^{ij}_q = \frac{e^2}{\beta}\sum_k \tr\left[\hat{G}_{k+\tfrac{q}{2}}\left(v^i_\vec{k}\hat{\tau}_0+v^i_S\hat{\tau}_3\right)\hat{G}_{k-\tfrac{q}{2}}\left(v^j_\vec{k}\hat{\tau}_0 + v^j_S\hat{\tau}_3\right) \right],
\end{equation}
where $\hat{G}_k = \left[(i\epsilon_n - \vec{k}\cdot\vec{v}_S)\hat{\tau}_0 - \xi_k^S\hat{\tau}_3 + \Delta\hat{\tau}_1\right]^{-1}$ is the Nambu Green's function. 
Unlike for the Bardasis-Schrieffer mode, here we keep the $q$ dependence of the Green's functions. 
Upon inserting resolutions of the identity to diagonalize the Green's function with the appropriate Bogoluibov transformation, $\hat{U}_\vec{k} = \begin{pmatrix}
u_\vec{k} & -v_\vec{k} \\
v_\vec{k} & u_\vec{k}
\end{pmatrix}$ with $u_\vec{k},v_\vec{k} = \sqrt{\frac{1}{2}\left(1 \pm \frac{\xi^S_k}{\lambda_k}\right)}$, and performing the Matsubara summation we have
\begin{multline} 
    \Pi^{ij}_q = e^2 \sum_{\vec{k}} \sum_{\alpha,\alpha'}\frac{n_F\left(E^{\alpha'}_{\vec{k}-\vec{q}/2}\right)-n_F\left(E^\alpha_{\vec{k}+\vec{q}/2}\right)}{i\Omega_m -\left(E^\alpha_{\vec{k}+\vec{q}/2} - E^{\alpha'}_{\vec{k}-\vec{q}/2}\right)} \left\lbrace v^i_\vec{k}v^j_\vec{k}\left(\ell_\vec{k,q}^2\delta_{\alpha,\alpha'} - p_\vec{k,q}^2\delta_{\alpha,-\alpha'}\right) + v^i_S v^j_S \left(n_\vec{k,q}^2 \delta_{\alpha,\alpha'} + m_\vec{k,q}^2 \ \delta_{\alpha,-\alpha'}\right) \right.\\
    \left. + \left(v^i_\vec{k} v^j_S + v^i_S v^j_\vec{k}\right) \ell_\vec{k,q}n_\vec{k,q}\ \alpha\, \delta_{\alpha,\alpha'} + \left(v^i_\vec{k} v^j_S - v^i_S v^j_\vec{k}\right) p_\vec{k,q}m_\vec{k,q}\  \alpha\,\delta_{\alpha,-\alpha'}\right\rbrace,
\end{multline}
where we have defined the superconductor coherence factors
\begin{gather}
    \ell_\vec{k,q} = u_+ u_- + v_+ v_- \qquad p_\vec{k,q} = u_+ v_- - v_+ u_-\\
    n_\vec{k,q} = u_+ u_- - v_+ v_- \qquad m_\vec{k,q} = u_+ v_- + v_+ u_-,
\end{gather}
using the shorthand notation for the Bogoliubov amplitudes $u_\pm = u_\vec{k\pm q/2}$ and similarly for $v_\pm$.

Analytic evaluation of this function keeping the full momentum and frequency dependence is unfeasible, so now we expand to first order in the deviation of the frequency from the cavity resonant frequency, $\delta\Omega = i\Omega - \omega_0$, which is the most that could be needed in the mode operator picture, and to second order in $|\vec{q}|$. 
Furthermore, we note that $v_\vec{k} \gg v_S$ and use this to make some further approximations, dropping terms with $v_S^2$ when there is a corresponding term appearing with $v_\vec{k}^2$.
We write the result of this expansion as
\begin{multline}
    \Pi^{ij}_q \approx x^{10,ij}_P(\phi_q)\,q \left(1-\frac{\delta\Omega}{\omega_0}\right) + x^{20,ij}_P(\phi_q)\,q^2 + \left(x^{00}_S + x^{01}_S\,\delta\Omega + x^{10}_S(\phi_q)\,q + x^{11}_S(\phi_q) \,q\,\delta\Omega + x^{20}_S(\phi_q)\,q^2\right) \delta_{ij}\delta_{i,\parallel} \\
    + \left[\left(x^{10,i}_{SPs}(\phi_q) + x^{10,i}_{SPa}(\phi_q)\right)\,q + \left(x^{11,i}_{SPs}(\phi_q) + x^{11,i}_{SPa}(\phi_q)\right)\,q\,\delta\Omega + \left(x^{20,i}_{SPs}(\phi_q) + x^{20,i}_{SPa}(\phi_q)\right)\,q^2\right]\delta_{j,\parallel} \\
    + \left[\left(x^{10,j}_{SPs}(\phi_q) - x^{10,j}_{SPa}(\phi_q)\right)\,q + \left(x^{11,j}_{SPs}(\phi_q) - x^{11,j}_{SPa}(\phi_q)\right)\,q\,\delta\Omega + \left(x^{20,j}_{SPs}(\phi_q) - x^{20,j}_{SPa}(\phi_q)\right)\,q^2\right]\delta_{i,\parallel}.
\end{multline}
The thirteen coefficients that appear in this expansion are given in Eq.~\eqref{eq:coeffs}.
Many of them are functions of the angle $\phi_q$, the angle vector $\vec{q}$ makes with axis defined by the supercurrent.
They are labeled with a subscript showing the type of vertices they arise from, $P$ for two paramagnetic vertices $\vec{v_k}$, $S$ for two supercurrent vertices $\vec{v}_S$, and $SP$ for one of each.
The secondary indices $s$ and $a$ label whether the term is symmetric or antisymmetric in $i,j$.
The superscript indices give the powers of $|\vec{q}|$ (first index) and $\delta\Omega$ (second index) that the coefficient multiplies. 
In these expressions we have used the shorthand notation $\delta n_\vec{k} = n_F(E^-_\vec{k}) - n_F(E^+_\vec{k})$, $\delta n''_\vec{k} = n''_F(E^-_\vec{k}) - n''_F(E^+_\vec{k})$ and $N'_\vec{k} = n'_F(E^+_\vec{k}) + n'_F(E^-_\vec{k})$.

\begin{subequations}\label{eq:coeffs}
\begin{gather}
    x^{10,ij}_P(\phi_q) = -e^2v_S \sum_\vec{k} \frac{1}{\omega_0} N'_\vec{k} v_\vec{k}^i v_\vec{k}^j \cos\phi_q \label{eq:firstcoeff}\\
    x^{20,ij}_{P}(\phi_q) = - \frac{e^2}{\omega_0^2} \sum_\vec{k} \left[\left(\frac{\xi_k^S}{\lambda_k}\right)^2 N'_\vec{k} + \frac{\Delta^2}{\lambda_k^3}\frac{\omega_0^2}{\omega_0^2-(2\lambda_k)^2}\delta n_\vec{k}\right] v_\vec{k}^i v_\vec{k}^j \ v_k^2\, \cos^2(\phi_k - \phi_q) \\
    x^{00}_S = 4e^2v_S^2 \sum_\vec{k} \frac{\Delta^2}{\lambda_k^2}\frac{\lambda_k}{\omega_0^2-(2\lambda_k)^2}\delta n_\vec{k}  \\
    x^{01}_S = -2e^2v_S^2\omega_0 \sum_\vec{k} \frac{\Delta^2}{\lambda_k^2} \frac{\lambda_k}{\left[\omega_0^2-(2\lambda_k)^2\right]^2}\delta n_\vec{k} \\
    x^{10}_S(\phi_q) = e^2 \omega_0 v_S^3 \sum_\vec{k} \frac{\Delta^2}{\lambda_k^2}\left[\frac{8\lambda_k}{\left[\omega_0^2-(2\lambda_k)^2\right]^2}\, \delta n_\vec{k} - \frac{1}{\omega_0^2-(2\lambda_k)^2} N'_\vec{k}\right] \cos\phi_q \\
    x^{11}_S(\phi_q) =  - e^2v_S^3 \sum_\vec{k} \frac{\Delta^2}{\lambda_k^2}\left[8\lambda_k\frac{3\omega_0^2 + (2\lambda_k)^2}{\left[\omega_0^2-(2\lambda_k)^2\right]^3}\,\delta n_\vec{k} - \frac{\omega_0^2+(2\lambda_k)^2}{\left[\omega_0^2-(2\lambda_k)^2\right]^2}N'_\vec{k}\right]\ \cos\phi_q \\
    x^{20}_S(\phi_q) = \frac{e^2 v_S^2}{2} \sum_\vec{k} \frac{\Delta^2}{\lambda_k^2}\left[\frac{ \Delta^2}{\lambda_k^3}\frac{\omega_0^2+(2\lambda_k)^2}{\left[\omega_0^2-(2\lambda_k)^2\right]^2} \delta n_\vec{k} - \left(\frac{\Delta}{\lambda_k}\right)^2\frac{N'_\vec{k}}{\omega_0^2-(2\lambda_k)^2} + \left(\frac{\xi_k^S}{\lambda_k}\right)^2 \frac{\lambda_k}{\omega_0^2-(2\lambda_k)^2} \delta n''_\vec{k}\right] v_k^2\,\cos^2(\phi_k-\phi_q) \\
    x^{10,i}_{SPs}(\phi_q) = - \frac{e^2 v_S}{\omega_0} \sum_\vec{k} \left(\frac{\xi_k^S}{\lambda_k}\right)^2N'_\vec{k} v_\vec{k}^i\ v_k\,\cos(\phi_k-\phi_q) \\
    x^{11,i}_{SPs}(\phi_q) = e^2v_S \sum_\vec{k} \left(\frac{\xi_k^S}{\lambda_k}\right)^2\frac{1}{\omega_0^2}N'_\vec{k} v_\vec{k}^i\ v_k\,\cos(\phi_k-\phi_q) \\
    x^{20,i}_{SPs}(\phi_q) = -e^2v_S \sum_\vec{k} \left(\frac{\xi_k^S}{\lambda_k}\right)^2\frac{2}{\omega_0^2}N'_\vec{k}v_\vec{k}^i\, v_k\, \cos(\phi_k-\phi_q) \cos\phi_q \\
    x^{10,i}_{SPa}(\phi_q) = \frac{e^2 v_S}{\omega_0} \sum_\vec{k}  \frac{\Delta^2}{\lambda_k^3} \frac{\omega_0^2}{\omega_0^2-(2\lambda_k)^2}\delta n_\vec{k}\ v_\vec{k}^i\ v_k\,\cos(\phi_k-\phi_q) \\
    x^{11,i}_{SPa}(\phi_q) = -e^2v_S \sum_\vec{k} \frac{\Delta^2}{\lambda_k^3}\frac{\omega_0^2+(2\lambda_k)^2}{\left[\omega_0^2-(2\lambda_k)^2\right]^2}\delta n_\vec{k}\ v_\vec{k}^i\ v_k\,\cos(\phi_k-\phi_q) \\
    x^{20,i}_{SPa}(\phi_q) = e^2 v_S \sum_\vec{k} \frac{\Delta^2}{\lambda_k^3}\left[\frac{\omega_0^2+(2\lambda_k)^2}{\left[\omega_0^2-(2\lambda_k)^2\right]^2}\delta n_\vec{k} - \frac{\lambda_k}{\omega_0^2-(2\lambda_k)^2}N'_\vec{k}\right]\,v_\vec{k}^i\, v_k\, \cos(\phi_k-\phi_q)\cos\phi_q \label{eq:lastcoeff}
\end{gather}
\end{subequations}

Because of the angular dependence inside the Fermi functions due to the Doppler shift in the energy, even the simplest of these expressions cannot be evaluated analytically.
After numerical evaluation and comparing the size of the terms in the expression for $\Pi^{ij}_q$, it so happens that only a single one of these terms is large enough to be of any importance--the constant $x^{00}_S$ term, which affects just the component of $\vec{A}_q$ parallel to the supercurrent.
This contribution to the photon action is then 
\begin{equation}
    S_\Pi = \frac{1}{2\beta}\sum_q x^{00}_S A^\parallel_{-q} A^\parallel_q = \frac{1}{2\beta}\sum_{q,\alpha,\beta} x^{00}_S \epsilon^\parallel_{\alpha,-\vec{q}} \epsilon^\parallel_{\beta,\vec{q}} A_{\alpha,-q} A_{\beta,q}.
\end{equation}

We now change to the mode basis using the transformation defined with the empty cavity part of the action, $A_{\alpha,q} = (a_{\alpha,q}+\bar{a}_{\alpha,-q})/\sqrt{2\omega_q}$.
After the usual approximation of discarding counterrotating terms ($\bar{a}\bar{a}$ and $aa$) the result is
\begin{equation}
    S_A\to S_a = \frac{1}{\beta}\sum_{q,\alpha,\beta} \bar{a}_{\alpha,q}\left[-i\Omega_m \delta_{\alpha\beta} + \omega_q \delta_{\alpha\beta} + x^{00}_S\frac{\epsilon^\parallel_{\alpha,-\vec{q}} \epsilon^\parallel_{\beta,\vec{q}}}{2\omega_q}\right]a_{\beta,q}.
\end{equation}
The last two terms comprise the effective photonic Hamiltonian in the polarization basis.
In our approximations the basis transformation induced by the polarization vectors is unitary up to an overall constant factor, so changing back from the polarization basis to the basis defined relative to the supercurrent direction diagonalizes the Hamiltonian,
\begin{equation}
    S_a = \frac{1}{\beta}\sum_{q} \left(\bar{a}^\parallel_q \bar{a}^\perp_q\right)\left[-i\Omega_m \hat{\mathbb{1}} + \begin{pmatrix} \omega_q + \Pi^S_q & 0 \\
    0 & \omega_q
    \end{pmatrix} \right]\begin{pmatrix} a^\parallel_q \\ a^\perp_q \end{pmatrix},
\end{equation}
where we define the only remaining part of the photonic self energy $\Pi^S_q = x^{00}_S/(L\,\omega_q)$.

\subsection{Coupling Term}
Finally we consider the coupling term in the action.
We replace $A_{\alpha,q}$ and $d_q$ with their definitions in terms of the mode operators $a_{\alpha,q}$ and $b_q$ and then perform the same transformation as above, from the polarization basis back to the Cartesian supercurrent basis. 
We then replace the imaginary frequency with the BS frequency, since that is the frequency at which the BS mode and photon bands hybridize. 
The result is
\begin{equation}
    S_{d-A}\to S_{b-a} = -\frac{i e v_S \Delta}{\beta}\sum_{\vec{k},q} \sqrt{\frac{2\,\Omega_\text{BS}}{L\,K\omega_q}} \frac{f_d(\phi_q)}{\lambda_k}\frac{\delta n_\vec{k}}{\Omega_\text{BS}^2 - (2\,\lambda_k)^2} \left(\bar{b}_q a^\parallel_q + \bar{a}^\parallel_q b_q\right),
\end{equation}
from which we then extract the coupling matrix element $g_q$ as in the main text. 
Altogether this gives the effective Hamiltonian for the coupled cavity photon-superconductor system,
\begin{equation}
    \check{H}^\text{eff}_q = \begin{pmatrix}
    \Omega_\text{BS} & g_q & 0 \\
    g_q & \omega_q + \Pi^S_q & 0 \\
    0 & 0 & \omega_q
    \end{pmatrix}.
\end{equation}
We see that this $3\times 3$ Hamiltonian decouples into a $2\times 2$ block and a single state. 
The block describes the hybridization of the BS mode with one photon mode, and the remaining state is the decoupled ``dark'' photon with the empty cavity dispersion $\omega_q$, which is unseen by the BS mode and is unaffected by the superconductor within our approximations. 
Since all $2\times 2$ matrices can be trivially diagonalized, the polariton dispersion can immediately be written
\begin{equation}
    E^{(\pm)}_q = \frac{\Omega_\text{BS} + \omega_q + \Pi^S_q}{2} \pm \sqrt{\left(\frac{\Omega_\text{BS}-\left(\omega_q+\Pi^S_q\right)}{2}\right)^2 + g_q^2}.
\end{equation}
These energies have corresponding eigenstates defined through
\begin{equation}
    \check{H}^\text{eff}_q \ket{E^{(\pm)}_q} = E^{(\pm)}_q\ket{E^{(\pm)}_q},
\end{equation}
which each have nontrivial overlap with both the uncoupled photon and BS states. 

\section{Methods for Numerical Solution}

We verified the results of our analytic model by numerically solving for the polariton dispersions.

The numerical method begins again with the effective Gaussian Matsubara action describing the coupled Bardasis-Schrieffer cavity-photon system.
Schematically this is
\begin{equation}
    S = -
    \frac{1}{2\beta}\sum_q\begin{pmatrix}
    d_{-q}&\mathbf{A}_{-q}
    \end{pmatrix}
    \begin{pmatrix}
    D_\text{BS}(q)^{-1}&\mathbf{g}(i\Omega_m)\\
    \mathbf{g}(-i\Omega_m)&\hat{D}_\text{phot}^{-1}
    \end{pmatrix}
    \begin{pmatrix}
    d_q\\
    \mathbf{A}_{q}
    \end{pmatrix},
\end{equation}
where the cavity propagator $\hat{D}^{-1} = \hat{D}_0^{-1} - \hat\Pi$ includes the self energy due to the superconductor.
At this stage the polariton modes can be found by solving of the frequency $z=i\Omega_m$ at which the inverse of the Green's function matrix vanishes.
To do so, we numerically solve for the roots of the determinant of the inverse Green's function $\det \hat{D}^{-1}(\Omega_{\mathbf{q}i}, \mathbf{q}) = 0$.
In particular the following algorithm was employed at each $\mathbf{q}$: noting that there are three roots that we are searching for
\begin{enumerate}
    \item An interval $[\omega_l, \omega_u]$ is chosen within which to search for solutions.
    \item An extremum $f$ of $\det \hat{D}^{-1}(\Omega, \mathbf{q})$ with respect to $\Omega$ is located by finding the roots of the first derivative with respect to $\Omega$ using the Newton-Raphson method in the vicinity of the Bardasis-Schrieffer frequency $\Omega_\text{BS}$.
    \item The other extremum is found by searching for the root of the first derivative in the interval $(\omega_l, f)$ or $(f, \omega_u)$ as determined by the sign of the function at the endpoints. This gives us two extrema $\{f_0, f_1\}$.
    \item Roots of $\det \hat{D}^{-1}(\Omega, \mathbf{q})$ are searched for using the Brent-Dekker method in the intervals $(\omega_l, f_0)$, $(f_0, f_1)$, and $(f_1, \omega_u)$
\end{enumerate}

For the second method, $d$ and $\mathbf{A}$ can be rewritten in terms of mode creation and annihilation operators to obtain an effective Hamiltonian as described in the main text.
The polariton dispersions are then simply obtained as eigenvalues of the Hamiltonian $\hat{H}(\mathbf q)$.

Comparing the results of the two methods shows that they are in excellent agreement as can be seen in the first figure of the main text.

Numerical integration and root-finding were performed using the GSL Scientific Library\cite{Galassi2009}.




\end{document}